\documentclass[%
 preprint,%
 amssymb, amsmath,%
 aip, jap,%
groupedaddress,
]{revtex4-1}

\usepackage{docs}%
\usepackage{bm}%
\usepackage[colorlinks=false,linkcolor=blue]{hyperref}%
\usepackage{float}
\usepackage{amsfonts,latexsym,eucal,color,graphicx,epsfig,amssymb,amsmath}
\expandafter\ifx\csname package@font\endcsname\relax\else
 \expandafter\expandafter
 \expandafter\usepackage
 \expandafter\expandafter
 \expandafter{\csname package@font\endcsname}%
\fi
\begin{document}

\title{A quantum interface to charged particles in a vacuum}%

\author{Hiroshi Okamoto}%
\email{okamoto@akita-pu.ac.jp}
\affiliation{Department of Electronics and Information Systems, Akita Prefectural University, Yurihonjo 015-0055, Japan}%

\date{May 2015}%

\begin{abstract}

A superconducting qubit device suitable for interacting with a flying electron has recently been proposed [H. Okamoto and Y. Nagatani, Appl. Phys. Lett. \textbf{104}, 062604 (2014)]. Either a clockwise or counter clockwise directed loop of half magnetic flux quantum encodes a qubit, which naturally interacts with any single charged particle with arbitrary kinetic energy. Here, the device's properties, sources of errors and possible applications are studied in detail. In particular, applications include detection of a charged particle without applying a force to it. Furthermore, quantum states can be transferred between an array of the proposed devices and the charged particle.

\end{abstract}

\maketitle

\section{Introduction}
 Quantum information processing methods may involve both the flying qubits and fixed qubits. One example is the quantum repeater \cite{quantum repeater}, where fixed qubits could be used to realize long-distance quantum communication based on flying photons. Distributed quantum computing is another example, which would also employ photons as flying qubits connecting different parts of the computer \cite{distQC,distQC exp}.

 Flying \emph{electrons}, combined with fixed superconducting qubits \cite{SC qubit review}, could also perform useful tasks. A known possible application of such a scheme is entanglement-assisted electron microscopy \cite{eeem, okamoto-nagatani} for radiation sensitive biological specimens \cite{Henderson review}. The use of a superconducting charge qubit \cite{eeem} and a radio frequency superconducting quantum interference device (rf-SQUID) qubit \cite{okamoto-nagatani} has been suggested to reduce the noise level down to the Heisenberg limit. In what follows, we will focus on the rf-SQUID qubit rather than the charge qubit because generally a magnetic qubit appears to be more convenient in practice, especially when high energy charged particles are used. In particular, we will find that the rf-SQUID qubit, or a variant of it, allows us to detect single-charged particles without applying a classical force on them. Charged particle detection with such a property could be useful in various relevant fields such as particle physics. More generally, we will show that bidirectional quantum information transfer between a single-charged particle and an array of qubits are possible in principle, opening ways to a wider range of applications. 

 The paper is organized as follows. 
 Section~\ref{sec: scheme} reviews the rf-SQUID qubit designed for our purpose. We then present how the single-charged particle detection works with our device. 
 In Section~\ref{sec: errors} we consider in detail various sources of charged particle detection errors. 
 In Section~\ref{sec: discussion}, after brief discussion of the operation sequence, we first discuss a possible application of the charged particle detector in biology. Second, to illustrate another more distant yet fundamentally sound possibility, we extend our single-qubit scheme to a multi-qubit scheme to enable bidirectional quantum information transfer between a charged particle and a superconducting quantum information processor. 

 We note the recent proposal on the use of \emph{trapped} electrons with superconducting qubits \cite{Daniilidis}. We assume that degrees of freedom other than the center-of-mass position of the charged particle, such as the spin degree of the electron, is well-isolated in our scheme and ignore such extra degrees of freedom in this paper. The symbol ``$e$'' denotes the positron charge.
 
\section{The main scheme}
\label{sec: scheme}
\begin{figure*}
\includegraphics[scale=0.2]{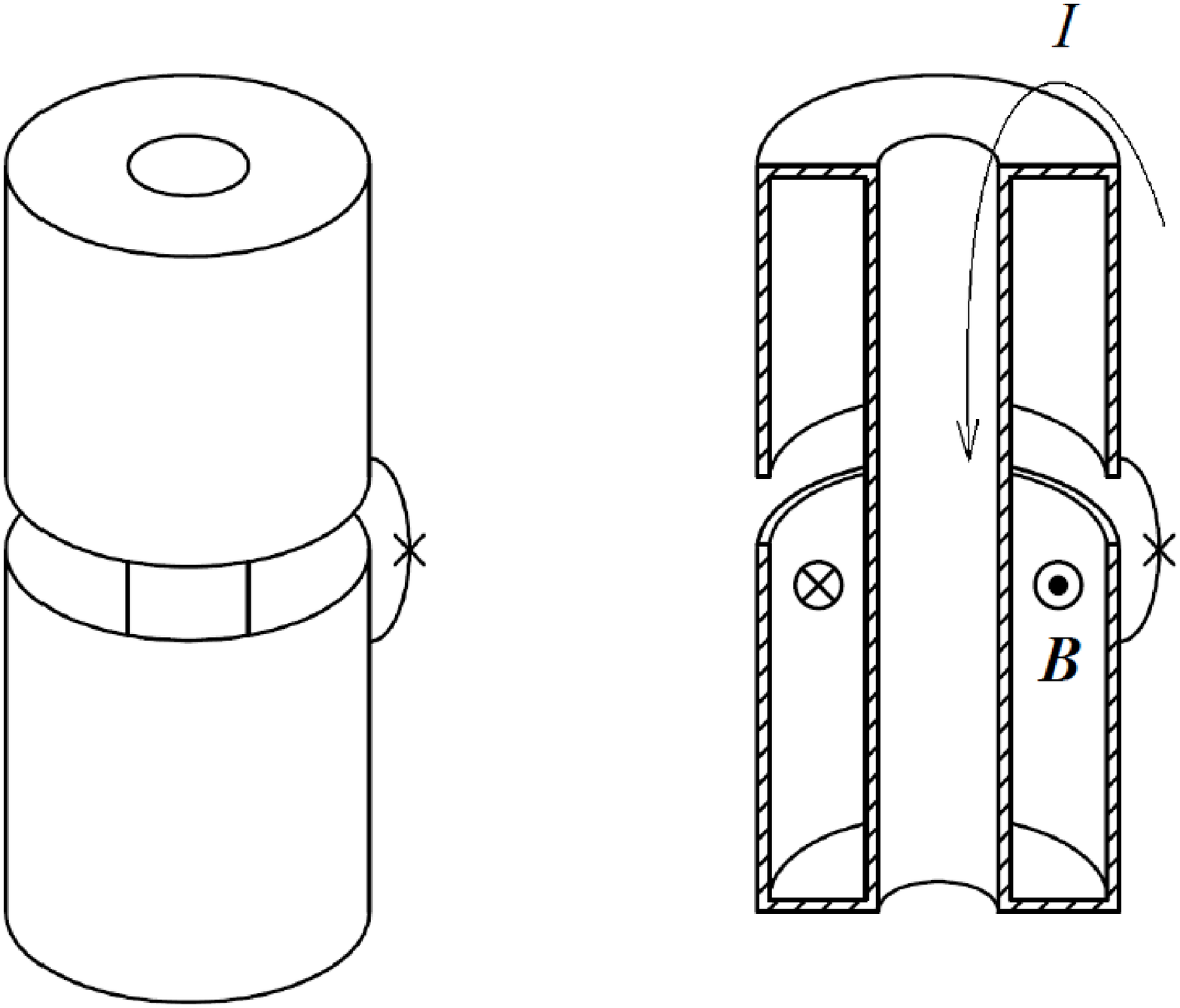}(a)\quad \quad \quad
\includegraphics[scale=0.2]{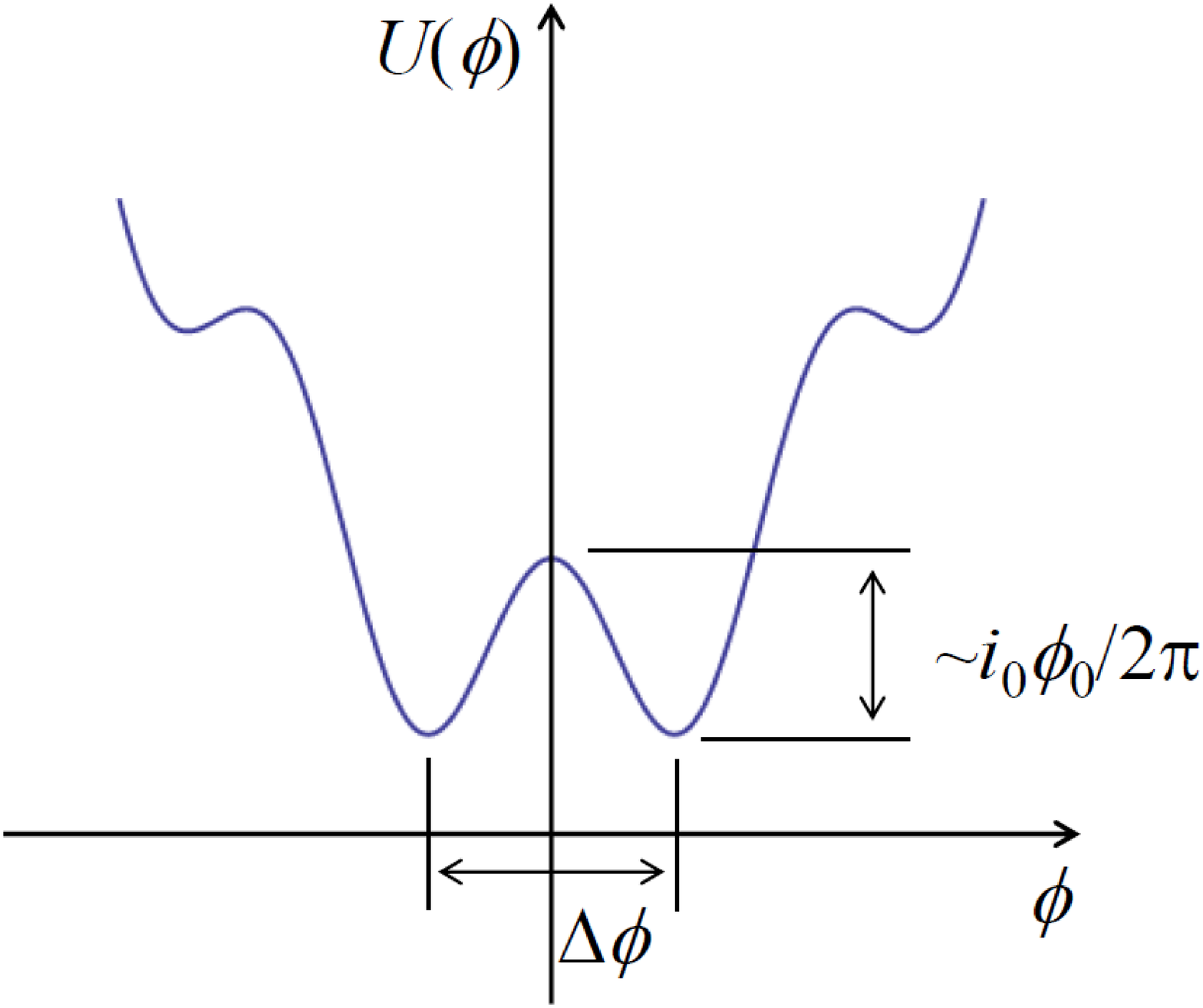}(b)\quad \quad \quad 
\includegraphics[scale=0.2]{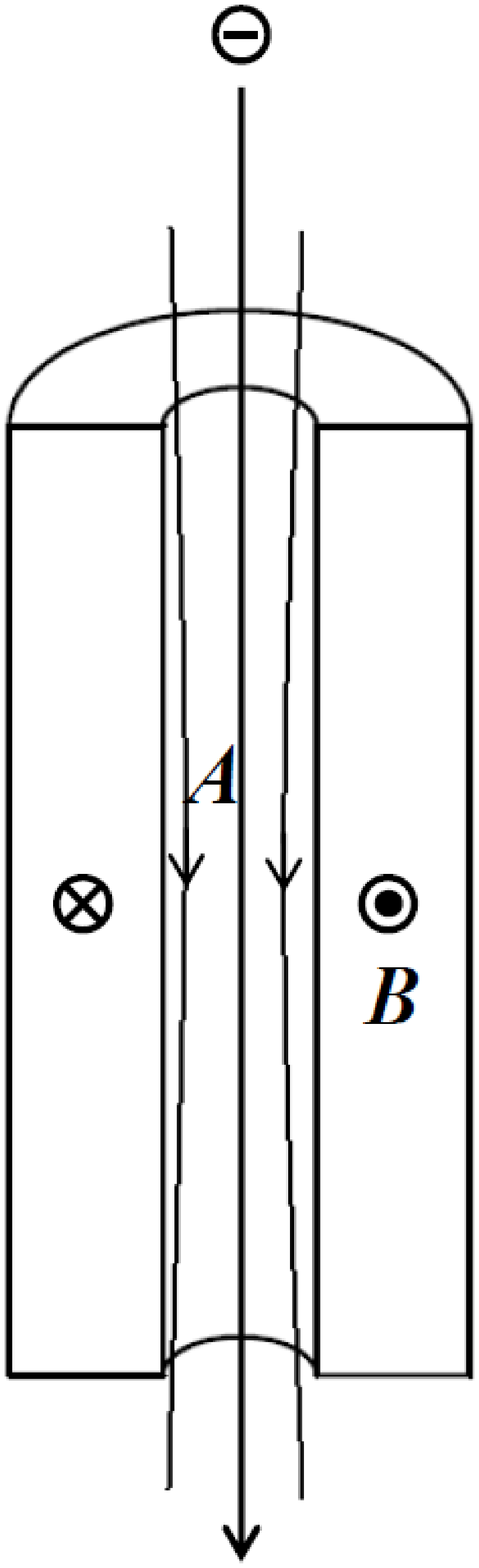}(c)

\caption{The rf-SQUID qubit. 
(a) The overall structure. A cross section is shown on the right. The persistent current flows through a double-walled tube, i.e. a hollow-ring with a slit. A Josephson junction (``X'' symbol) is inserted across the slit. A ring-shaped magnetic flux is trapped inside the hollow ring. This particular state of the flux represents the state $|0\rangle_{q}$.
(b) The potential energy landscape of an rf-SQUID that is biased with an external magnetic flux $\phi_{0}/2$.
(c) A negatively single-charged particle such as the electron, represented by a particle with a negative sign, follows the flow of the $\boldsymbol{A}$-field corresponding the state $|0\rangle_{q}$.}
\label{fig:fig1}
\end{figure*}

 Figure~\ref{fig:fig1} (a) shows a qubit based on the rf-SQUID \cite{SQUID qubit experiment}, where the superconductor has hollow ring geometry so that all the associated magnetic flux is essentially within the device \cite{okamoto-nagatani}. Henceforth, the term ``rf-SQUID'' refers to this type of device in this paper.
 The idea is to produce a ring of half magnetic flux quantum hovering in the vacuum, which is in a quantum mechanically superposed state of two opposing directions of the magnetic flux. By the Aharonov-Bohm (AB) effect \cite{AB effect}, the single-charged matter wave passing through the ring acquires a phase shift relative to the wave passing outside the ring, without ever receiving a force unless the wave directly hits the ring. We defer to a future study an important aspect of how to realize such a hovering magnetic flux ring in practice, while noting that a preliminary consideration has already appeared \cite{okamoto-nagatani}. It is possible that the eventual implementation of our scheme would involve a superconducting device that is not precisely an rf-SQUID qubit, for technical reasons that will be discussed briefly in Sec.~\ref{operation sequence}. However, having a ring of half magnetic flux quantum in the matter wave of a charged particle is an essential part of our scheme.

 To study dynamics of rf-SQUID from the circuit perspective, we introduce the inductance of the SQUID loop $L$, the critical current of the Josephson junction $i_{0}$ and the junction capacitance $C$. The rf-SQUID loop is magnetic-flux-biased with a half magnetic flux quantum $\phi_{0}/2=h/4e$. The potential energy of the rf-SQUID is 
\begin{equation}
\label{eq: squid potential}
U\left(\phi\right)=\frac{\phi^{2}}{2L}-E_{J}\cos\left(\frac{2\pi\phi}{\phi_{0}}+\pi\right),
\end{equation}
where $E_{J}=i_{0}\phi_{0}/2\pi$ (See Fig.~\ref{fig:fig1} (b)). Hence, the two qubit states $|0\rangle_{q}$ and $|1\rangle_{q}$, respectively associated with the trapped flux $\phi \approx -\phi_{0}/2$ and $\phi_{0}/2$, have the same energy. The subscript ``{\it q}'' stands for ``qubit''. We also define $|s\rangle_{q}=\left(|0\rangle_{q}+|1\rangle_{q}\right)/\sqrt{2}$ and $|a\rangle_{q}=\left(|0\rangle_{q}-|1\rangle_{q}\right)/\sqrt{2}$ for later use. For simplicity, we first assume that these values are exactly $\pm \phi_{0}/2$ and quantum mechanically well-defined. 

 We use a few conventions when studying the device from the electromagnetic perspective. We will use the Coulomb gauge $\mathrm{div}\boldsymbol{A}=0$ for the vector potential $\boldsymbol{A}$ throughout the paper, and assume that $\boldsymbol{A}$ goes to zero as the distance from the ring goes to infinity. To have a simple $\boldsymbol{A}$-field structure, we further assume that the magnetic field is zero outside the qubit device. The structure of the qubit is intended to shift the phase of a charged particle wave passing through the hollow ring by a phase angle $\pm\pi/2$ depending on the qubit state, while not much affecting the phase of the charged particle waves passing outside the ring. This statement makes sense because we fixed the gauge and stipulated several additional conditions.

 Note that this type of interaction remains the same for any single-charged particle, i.e. a particle having a charge $\pm e$, regardless of the species and kinetic energy of the particle. Without loss of generality, throughout the paper we will assume that the charged particle has a negative charge $-e$. Let the quantum state of the matter wave of the charged particle passing through the hollow ring be $|a\rangle=\left(|0\rangle-|1\rangle\right)/\sqrt{2}$ and the wave passing outside the ring be $|s\rangle=\left(|0\rangle+|1\rangle\right)/\sqrt{2}$, 
where we also introduced the states $|0\rangle,|1\rangle$. Without loss of generality, we can state that the charged particle state $|a\rangle$ transforms to $-i|a\rangle$ when the rf-SQUID qubit is in the state $|0\rangle_{q}$ and likewise the state transforms to $i|a\rangle$ for the qubit state $|1\rangle_{q}$, while the state $|s\rangle$ remains the same. (Since the charge is negative, $\hbar \boldsymbol{k} + e \boldsymbol{A}$ remains constant, where $\boldsymbol{k}$ denotes the wave vector of the wavefunction $\psi  \propto e^{i \boldsymbol{k} \cdot \boldsymbol{r}}$ of the charged particle. See Figs. 1 (a) and 1 (c).)
 We assume that there is zero or only negligible amount of the matter wave component impinging on the body of the hollow ring. This may be done deliberately in certain cases, for example by placing a suitable stencil mask at the upstream of the charged particle beam \cite{okamoto-nagatani}.
 Now, consider a process comprising 2 steps. First, the state $|a\rangle$, and not the state $|s\rangle$, receives a phase shift $\pi/2$ and becomes $i|a\rangle$ by a classical charged-particle optical component, which we will call a $\pi/2$ phase shifter \cite{Pi/2 EM phase plate}. Second, the matter wave passes the rf-SQUID. This whole process flips the sign of the state $|a\rangle$ if and only if the rf-SQUID qubit is in the state $|1\rangle_{q}$. Hence, the process represents a controlled-not (CNOT) gate \cite{okamoto-nagatani}, where the rf-SQUID qubit with its logical value represented by the basis states $\left\{ |0\rangle_{q},|1\rangle_{q}\right\} $ controls the value associated with the charged particle in terms of the basis states $\left\{ |0\rangle,|1\rangle\right\} $.

The two qubits of any quantum CNOT gate swap their roles of either controlling or being controlled, upon basis change by the Hadamard transform \cite{nielsen-chuang}. In our case, the state of the rf-SQUID qubit in terms of the states $\left\{ |s\rangle_{q},|a\rangle_{q}\right\} $ is flipped if and only if the matter wave is in the state $|a\rangle$. This immediately suggests a use of the rf-SQUID as a non-destructive charged particle counter, because its quantum state flips if and only if a charged particle flies through it. 

To evaluate the above crude idea more quantitatively, we need more detailed understanding of its physics. 
 Figure 1 (c) shows a ring-shaped magnetic flux $\boldsymbol{B}$ associated with the state $|0\rangle_{q}$, around which the vector potential $\boldsymbol{A}$ swirls. The roles played by current density $\boldsymbol{j}$ and magnetic flux $\boldsymbol{B}$ in an ordinary solenoid are instead
played respectively by $\boldsymbol{B}$ and $\boldsymbol{A}$ in our device. To the first approximation, appreciable vector potential exists only inside the bore of the rf-SQUID. Let us consider how the charged particle in the state $|a\rangle$, flying through the qubit, flips the qubit state.
 First, the initial qubit state is $|s\rangle_{q}=\left(|0\rangle_{q}+|1\rangle_{q}\right)/\sqrt{2}$. The initial state of the charged particle is $|a\rangle$, and the $\pi/2$ phase shifter described above may as well be absent because it would change nothing except the overall phase. Thus, the initial state of the combined system is 
\begin{equation}
|\psi_{0}\rangle=|a\rangle|s\rangle_{q}=\frac{|a\rangle|0\rangle_{q}+|a\rangle|1\rangle_{q}}{\sqrt{2}}.
\end{equation}
 Upon interaction with the qubit, the charged particle wave follows or goes against the $\boldsymbol{A}$-field, resulting in a phase shift $\pm\pi/2$.
 The resultant state of the combined system after interaction is 
\begin{equation} \label{eq: entangle1}
\frac{e^{-i\pi/2}|a\rangle|0\rangle_{q}+e^{i\pi/2}|a\rangle|1\rangle_{q}}{\sqrt{2}},
\end{equation}
which equals, up to an overall phase factor, 
\begin{equation}
|\psi_{1}\rangle=\frac{|a\rangle|0\rangle_{q}-|a\rangle|1\rangle_{q}}{\sqrt{2}}.
\end{equation}
 Passing the phase factor to the qubit, the charged particle stays in the initial state as $|\psi_{1}\rangle=|a\rangle\left(|0\rangle_{q}-|1\rangle_{q}\right)/\sqrt{2}=|a\rangle|a\rangle_{q}$.

\section{Evaluation of errors}
\label{sec: errors}

\subsection{Vector potential outside the rf-SQUID}
\label{subsec: A outside the device}
 The first cause of error that we consider with respect to charged particle detection is non-ideal $\boldsymbol{A}$-field distribution. Specifically, the difference $\Delta\theta$ between the two phase shifts associated with two qubit states $|0\rangle_{q},|1\rangle_{q}$ is generally less than $\pi$ because of the $\boldsymbol{A}$-field outside the bore of the rf-SQUID. In contrast, the experiment demonstrating the AB effect \cite{tonomura ab experiment} measures the phase shift difference between two paths, which together encircles the magnetic flux.
 Writing $\Delta\theta=\pi-\delta$, equation~\eqref{eq: entangle1} is modified to be
\begin{equation}
\frac{e^{-i  \frac{(\pi-\delta )}{2}}|a\rangle|0\rangle_{q}+ e^{i  \frac{(\pi-\delta )}{2}} |a\rangle|1\rangle_{q}}{\sqrt{2}}
\cong -i|a\rangle|a\rangle_{q} + \frac{\delta }{2}|a\rangle|s\rangle_{q},
\end{equation}
where higher order terms in $\delta$ are ignored. Hence, the error probability for charged particle detection is $\approx\delta^{2}/4$ if $\delta\ll1$. To estimate $\delta$, let the length of the inner bore of the rf-SQUID be $l$ and the radius of the bore be $r\ll l$. Let the magnitude of the $\boldsymbol{A}$-field inside the bore be $A_{bore}$, which we assume to be highly uniform. In other words, just as a good solenoid generates a highly uniform magnetic field $\boldsymbol{B}$ from an electric current, so does our rf-SQUID generate a highly uniform $\boldsymbol{A}$-field from the magnetic flux. Furthermore, as the operating solenoid may be seen as producing two `magnetic charges' of opposite signs placed at both the ends, the rf-SQUID may be seen as holding a pair of `vector potential charges' (VPCs) $q_{A}=\pm\pi r^{2}A_{bore}$ at the two ends, disregarding the $\boldsymbol{A}$-field inside the bore. Treating these two VPCs as well-separated `point charges', $\boldsymbol{A}$-field distribution outside the rf-SQUID can be computed in the same manner the $\boldsymbol{E}$-field is computed in electrostatics. Specifically, since we disregard the $\boldsymbol{A}$-field inside the bore in this particular consideration, we can define a `potential' $\varphi_{A}$ that satisfies $\boldsymbol{A}=-\mathrm{grad}\varphi_{A}$. In particular, the `potential' near one of the VPCs, ignoring the influence of another, is given as $\varphi_{A} \approx q_{A}/4\pi r$, where $r$ is the distance between the measuring point and the VPC. The potential difference $\Delta\varphi_{A}$ between the two VPCs would be infinity if these were indeed point charges, but the VPCs have the `size' $\approx r$. Hence the potential difference is approximately $\Delta\varphi_{A} \approx q_{A}/2\pi r\approx rA_{bore}/2$, again ignoring the influence of the distant VPC when evaluating the `potential' at the either end of the rf-SQUID.
 The integral of $\boldsymbol{A}$ along a closed path $C$ interlinking the rf-SQUID once should satisfy 
\begin{equation}
\oint_{C}\boldsymbol{A}\cdot d\boldsymbol{l}\approx A_{bore}\left(l+\frac{r}{2}\right) = \frac{\phi_{0}}{2}. 
\end{equation}
Since the charged particle flies only inside the bore, the error is $\delta\approx r/2l$. This suggests that an aspect ratio of $l/r\approx10$ would be sufficient to achieve $\lesssim1\mathrm{\%}$ error or even less for this particular error source. Finally, we remark that the error discussed here does not affect the performance of entanglement-enhanced electron microscopy \cite{eeem, okamoto-nagatani}, where the phase difference between two paths matter.

\subsection{Shift in the minima of the potential landscape}
The second source of phase error is the non-ideal amount of magnetic flux inside the rf-SQUID. The separation $\Delta\phi$ between the two potential minima shown in Fig. \ref{fig:fig1} (b) is ideally $\phi_{0}$, but is somewhat less than that. On the other hand, the difference of the phase shift experienced by the flying charged particle in the state $|a\rangle$, with respect to the two states of the rf-SQUID $|0\rangle_{q}$ and $|1\rangle_{q}$, is ideally $\pi$ but is somewhat smaller $\pi - \varepsilon$. Hence we write 
\begin{equation}
\label{eq: epsilon}
\pi \Delta\phi/\phi_{0} = \pi - \varepsilon.
\end{equation}
 The error probability in terms of charged particle detection is $\approx\varepsilon^{2}/4$ by the same reasoning used in Sec. \ref{subsec: A outside the device}. On the other hand, the position of the potential minima in Fig. \ref{fig:fig1} (b) equals $\phi=\Delta\phi/2 = \left(\phi_{0}/2\right)\left(1-\varepsilon/\pi\right)$. This should satisfy $dU\left(\phi\right)/dt=0$, where $U\left(\phi\right)$ is given in eq. \eqref{eq: squid potential}. Hence $\pi-\varepsilon=\beta\sin\varepsilon$ follows, where $\beta/2\pi=Li_{0}/\phi_{0}$. Numerical calculation reveals that $Li_{0}\gtrsim 2.4\phi_{0}$  should be satisfied to have $\lesssim1\mathrm{\%}$ detection error.

\subsection{Excitation of the qubit state}
\label{Excitation of the qubit state}
The flux qubit is not strictly discrete in the sense that spin $1/2$ is discrete, and therein lies another source of an error. For example, the wavefunction $\psi_{q}\left(\phi\right)= ~_{q}\langle\phi|0\rangle_{q}$ of the rf-SQUID, where $|\phi\rangle_{q}$ is an eigenstate of the magnetic flux $\phi$, is not fully localized at the potential minimum at $\phi=\pm\left(1/2-\varepsilon/2\pi\right)\phi_{0}$. To see the effect of it, notice a relation $|0\rangle_{q}=\int\psi_{q}\left(\phi\right)|\phi\rangle_{q}d\phi$ for the basis system $\left\{ |\phi\rangle_{q}\right\} $ normalized as $~_{q}\langle \phi|\phi'\rangle_{q}=\delta(\phi-\phi')$.
 Since the state $|\phi\rangle_{q}$ induces a phase shift $\pi\phi/\phi_{0}$ to the charged particle wave, after interaction with the charged particle the qubit is left in the state $|0'\rangle_{q}=\int\psi_{q}\left(\phi\right)e^{i\pi\phi/\phi_{0}}|\phi\rangle_{q}d\phi$. This state is no longer exactly $|0\rangle_{q}$, meaning that there is a finite probability $p_{l}=1-\left|~_{q}\langle0'|0\rangle_{q}\right|^{2}$ that the qubit state is leaked out of its ``logical'' Hilbert space spanned by $|0\rangle_{q}$ and $|1\rangle_{q}$. To estimate the wavefunction spread, first use the standard method to obtain the Hamiltonian $H=q^{2}/2C+\phi^{2}/2L+E_{J}\cos\left(2\pi\phi/\phi_{0}\right)$ and the commutation relation $\left[\phi,q\right]=i\hbar$. To focus on one of the two potential minima, consider a purely harmonic potential that fits one of the two minima of $U\left(\phi\right)$.
Differentiating $U\left(\phi\right)$ twice, the effective inductance $\left(d^{2}U\left(\phi\right)/d\phi^{2}\right)^{-1}$ at the potential minimum $\phi=\left(1/2-\varepsilon/2\pi\right)\phi_{0}$ is 
\begin{equation}
L_{e}\approx \frac{LL_{J}}{L\left(1-\varepsilon^{2}/2\right)+L_{J}}= \frac{\beta}{\beta\left(1-\varepsilon^{2}/2\right)+1} L_{J}\approx\frac{\beta}{\beta+1}L_{J},
\end{equation}
where $L_{J}=\phi_{0}/2\pi i_{0}=L/\beta$ is the Josephson inductance at zero phase difference across it. Using this, the original Hamiltonian is approximated with a Hamiltonian of a harmonic oscillator, i.e. $H'=q^{2}/2C+\left(\phi-\phi_{0}/2\right)^{2}/2L_{e}$.
 The ground state is 
\begin{equation}
\psi_{q}\left(\phi\right)=\frac{1}{\sqrt[4]{\pi\phi_{1}^{2}}}e^{-\left(\phi-\phi_{0}/2\right)^{2}/2\phi_{1}^{2}},
\end{equation}
where $\phi_{1}^{2}=\hbar\sqrt{L_{e}/C}$. Hence we obtain $\left|~_{q}\langle0'|0\rangle_{q}\right|^{2}$ as
\begin{equation}
\left|\int_{-\infty}^{\infty}d\phi\left|\psi_{q}\left(\phi\right)\right|^{2}\cos\frac{\pi\left(\phi-\phi_{0}/2\right)}{\phi_{0}}\right|^{2}=e^{-\frac{\pi^{2}}{2}\left(\frac{\phi_{1}}{\phi_{0}}\right)^{2}}
\end{equation}
and the leakage probability is $p_{l}\approx\left(\pi^{2}/2\right)\left(\phi_{1}/\phi_{0}\right)^{2}$, or 
\begin{equation}
p_{l}\approx\sqrt{\frac{\beta}{\beta+1}}\sqrt{\frac{E_{C}}{8E_{J}}}
\end{equation}
in terms of $E_{J}$ and $E_{C}=e^{2}/2C$. Assuming $\sqrt{\beta/\left(\beta+1\right)}\approx1$, the ratio $E_{C}/E_{J}$ should be $\approx10^{-3}$ to achieve $\approx 1\mathrm{\%}$
detection error, which is not unusual for an rf-SQUID qubit \cite{SQUID qubit experiment}.

\subsection{Backaction to the flying charged particle}
 As discussed earlier, in charged particle detection, the charged particle state $|a\rangle$ is fully disentangled from the rf-SQUID qubit state and remains in the initial state after going through the rf-SQUID. Hence, neither spatial nor temporal coherence of the charged particle wave is important. 
 This remarkable robustness suggests that a classical charged particle suffices to explain the device operation. Figure 2 (a) shows a cross section of an rf-SQUID, in which a negative charged particle goes through. The rf-SQUID is electrically grounded to a conducting wall with a wire. The charged particle induces the opposite positive charge on the surface of nearby conductors. The positive charge moves along as the charged particle goes through the hollow ring. The reader will see that a total of charge $e$ will flow through the Josephson junction from left to right, especially when the hollow ring is such that all the electric field lines from the charged particle terminates on the surface of the rf-SQUID at one point of time during the charged particle's passage. (Here we do not pursue the case of an electrically floating rf-SQUID. While the conducting wire can clearly be replaced by a large capacitor, a full analysis remains to be done.)

Note that the current generated by movement of the induced charge flows on the outer surface of the rf-SQUID comprising finite-thickness superconducting films, whereas the current keeping the magnetic flux inside the hollow ring flows on the inner surface of the device. Since these two currents meet essentially only at the Josephson junction, we model the rf-SQUID circuit as in Fig. 2 (b), where a current source, producing a current that integrates to $e$, is attached to near the both sides of the Josephson junction. To make the analysis easier, we replace the current source by a large inductor $L_{L}$ that traps a large magnetic flux $\phi_{L}$, generating a bias current $i_{b}=\phi_{L}/L_{L}$. We can do this because the rf-SQUID does not ``care'' about what kind of current source is used. (See Fig. 2 (c). This inductor-based current source is only for our thought experiment and need not exist. One is free to imagine changing $i_{b}$ at will by mechanically deforming the inductor $L_{L}$, for example.) Since the total magnetic flux $\phi_{t}=\phi+\phi_{L}$, in addition to the bias flux $\phi_{0}/2$, is firmly trapped within a superconductor, it is a constant. Hence, the potential energy of the system is 
\begin{equation}
U'\left(\phi\right)=\frac{\phi^{2}}{2L}+\frac{\left(\phi_{t}-\phi\right)^{2}}{2L_{L}}-E_{J}\cos\left(\frac{2\pi\phi}{\phi_{0}}+\pi\right)\approx\frac{\phi^{2}}{2L}-i_{b}\phi+E_{J}\cos\left(\frac{2\pi\phi}{\phi_{0}}\right)+\mathrm{const.},
\end{equation}
where we used $\phi_{t}\approx\phi_{L}\gg\phi$ and assumed the absence of mutual inductance between $L$ and $L_{L}$. (Incidentally, we found that the bias flux $\phi_{0}/2$ may be applied by this type of current biasing. However, the required bandwidth of such a bias line may well entail a noise current directly fed to the Josephson junction, especially when the control signal comes from an external circuit.) We set $i_{b}=e/T$ for a time duration $T$. Then, one potential minimum goes up by an energy amount $\Delta E=i_{b}\phi_{0}/2=h/4T$, whereas the other minimum goes down by the same amount. As expected, this results in the phase difference $2\Delta ET/\hbar=\pi$ between the states $|0\rangle_{q},|1\rangle_{q}$ because this is the difference in action at least for sudden or adiabatic changes of the potential.
\begin{figure*}
\includegraphics[scale=0.25]{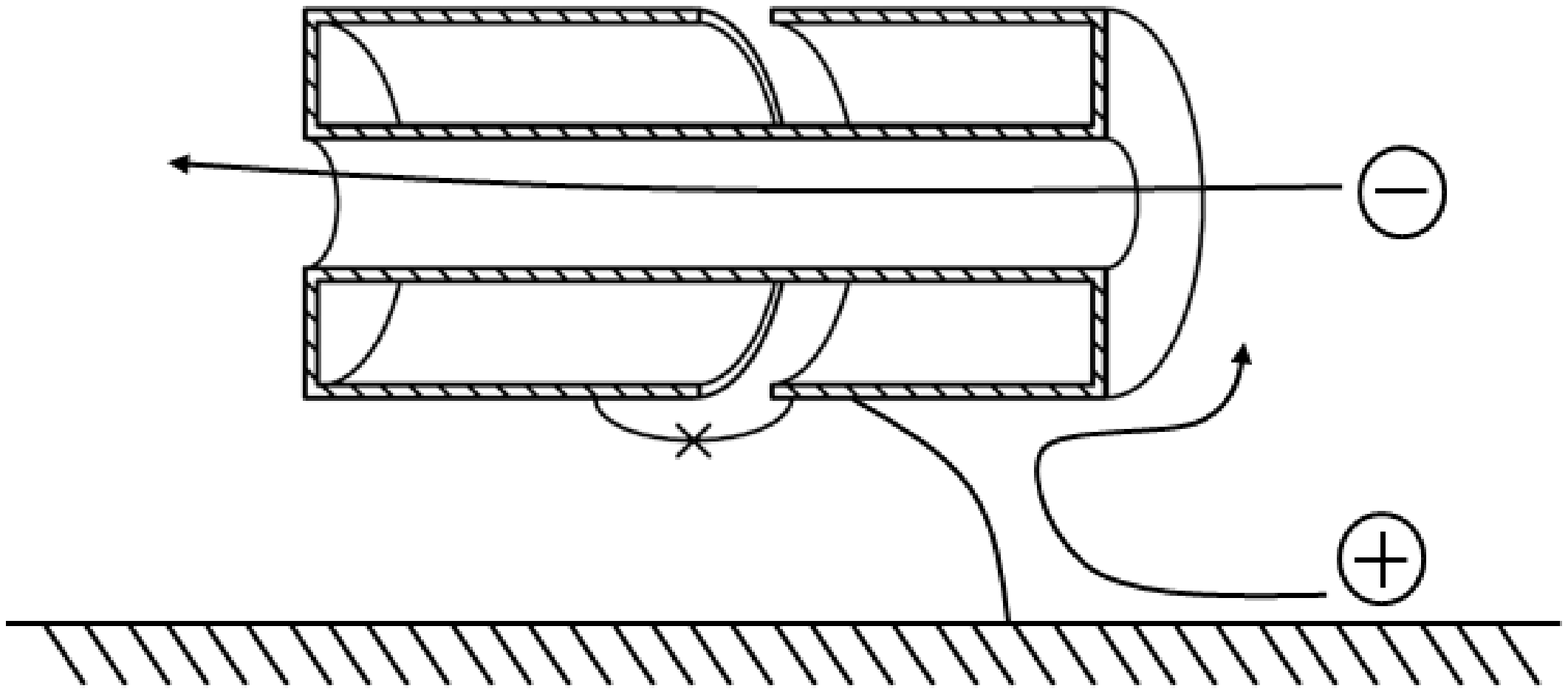}(a)\quad \quad 
\includegraphics[scale=0.25]{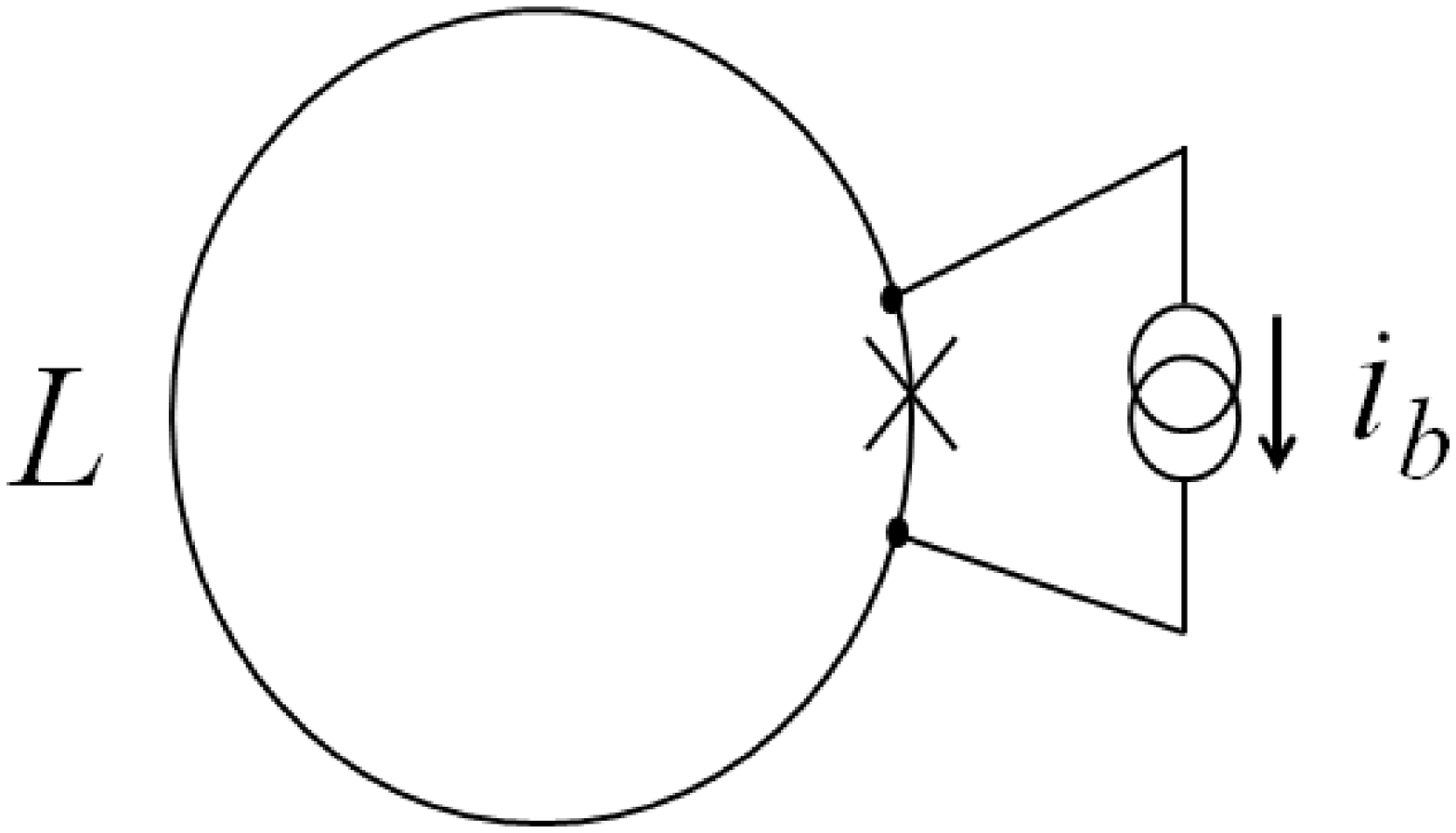}(b)\quad \quad 
\includegraphics[scale=0.25]{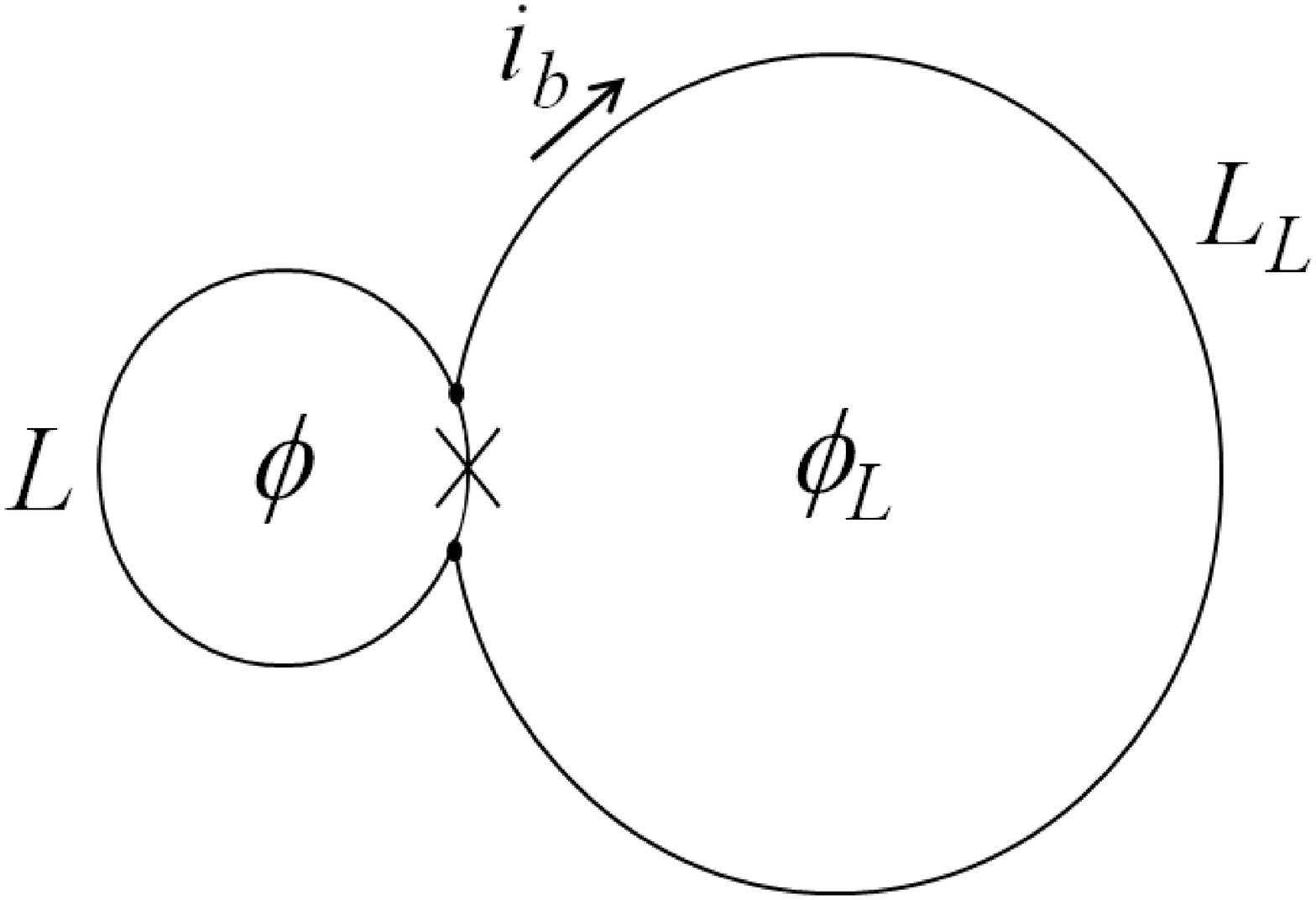}(c)

\caption{Non-destructive charge counting operation. 
(a) Induced positive charge flows on conductor surfaces as the charged particle flies through the rf-SQUID. 
(b) The flying charged particle is modeled as a current source, which in turn is modeled as another superconducting ring as in (c).}
\end{figure*}

The above argument, although not rigorous, is useful because it provides insights into `backaction' to the charged particle in our device. We consider two mechanisms. The first mechanism is due to the motion of the induced positive charge on the inductive surface of the rf-SQUID. This could influence the motion of the charged particle in essentially a non-dissipative way because the induced charge has ``inertia'' due to the inductance. The dissipative part due to the possible presence of quasiparticles will be considered in the next section. This mechanism is generic and similar effect should be present in any conductor \cite{hasselbach dissipation}. 

The second, perhaps more interesting, mechanism transfers energy from the charged particle to the rf-SQUID. In a sense we already know this because the state of the rf-SQUID evolves from $|s\rangle_{q}$ to $|a\rangle_{q}$ as the charged particle goes through the device. These states have different energy due to energy level splitting because $|0\rangle_{q}$ and $|1\rangle_{q}$ are coupled through the tunneling barrier in $U\left(\phi\right)$. (Here we put aside the possibility of excitation within a single potential well around one of the potential minima, which we have already discussed in Sec. \ref{Excitation of the qubit state}.) However, it is useful to examine this process from the present half-classical perspective, because we want to know exactly how the charged particle is decelerated, although an electrostatic force must be responsible because of lack of other candidates. Initially, the rf-SQUID is in the state $\left( |0\rangle_{q}+|1\rangle_{q} \right)/\sqrt{2}$. When the charged particle goes half way through the rf-SQUID, after factoring out the charged particle state, the state of the rf-SQUID is
\begin{equation}
\label{phase evolution}
\frac{e^{-i \delta}|0\rangle_{q}+e^{i \delta}|1\rangle_{q}}{\sqrt{2}},
\end{equation}
where $\delta$ is a small positive real number increasing with time. Although we have essentially seen this in Eq. \eqref{eq: entangle1}, below we consider this from the perspective centered on energetics of the rf-SQUID. The induced charge motion tends to force magnetic flux, opposite to that generated by the $|0\rangle_{q}$ state, into the rf-SQUID inductance ring (See Figs. 1 (a) and 2 (c). Suppose that the current in the inductor $L$ in Fig. 2 (c) flows counterclockwise for the state $|0\rangle_{q}$ to produce the magnetic flux $\phi$, in accordance with the right hand side of Fig. 1 (a). The current $i_{b}$ in Fig. 2 (c) generates a magnetic flux $\phi_{L}$ in $L_{L}$ with the direction opposite to $\phi$. Note that $i_{b}$ does not contribute to $\phi$ because of the absence of mutual inductance. The finite amount of flux $\phi_{L}$ ``wants'' to go into the loop $L$ unless the inductance $L_{L}$ is infinitely large.). Arguably, this makes the energy of the $|0\rangle_{q}$ state higher than that of $|1\rangle_{q}$, resulting in the phase factor in Eq. \eqref{phase evolution} because of the $e^{-iEt/\hbar}$ dependence on time. This is valid at least in the following two limits: the energy level actually goes up in the adiabatic approximation, and the potential energy term contributes to the above phase evolution in the sudden approximation. 

At some moment the state is $\approx \left( |0\rangle_{q}+i|1\rangle_{q}\right) / \sqrt{2}$ up to the overall phase factor, which has an associated charge, albeit ill defined, because the charge operator is $q = -i\hbar\partial/\partial\phi$. (Since the phase of the rf-SQUID wavefunction $\psi_{q}\left(\phi\right)$ rotates mostly within the tunnel barrier, the associated probability amplitude is small when $E_{J}$, or the height of the tunnel barrier, is large.) In order to determine the polarity of the Josephson junction charging, assume for the moment the state is exactly 
\begin{equation}
\frac{|0\rangle_{q}+i|1\rangle_{q}}{\sqrt{2}} \propto \frac{|s\rangle_{q}-i|a\rangle_{q}}{\sqrt{2}},
\end{equation}
and is freely evolving in the unbiased ($i_{b}=0$) potential. Since the state $|a\rangle_{q}$ has a larger energy than the state $|s\rangle_{q}$, this state is on the path evolving from $\left( |s\rangle_{q}+|a\rangle_{q} \right) /\sqrt{2} = |0\rangle_{q}$ to $\left( |s\rangle_{q}-|a\rangle_{q} \right) /\sqrt{2} = |1\rangle_{q}$. From the direction of the current shown in Fig. 1 (a) for the state $|0\rangle_{q}$, we find that the lower electrode of the Josephson junction is positively charged when the state is $\left( |s\rangle_{q}-i|a\rangle_{q} \right) /\sqrt{2}$. Thus, it takes positive work to supply the positive charge to the Josephson junction. This in turn means that the negative charged particle receives a `pulling' force from the positive induced charge, which moves with dissipation on the SQUID surface.

It is possible to make the above argument somewhat more quantitative. The amount of the charge $q$ on the Josephson junction is such that $q/C$ equals the electromotive force, which has the dimension of electrostatic potential, generated by the time-varying magnetic flux inside the SQUID. The oscillation frequency $\omega$ between the two states $|0\rangle_{q},|1\rangle_{q}$, which respectively corresponds to the two directions of the magnetic flux, represents the energy splitting $\hbar \omega$ between the states $|s\rangle_{q}$ and $|a\rangle_{q}$. On the other hand, the electromotive force is $\sim 2\phi_{0}\omega/2\pi \sim \sim \hbar \omega / e$. The first factor 2 comes from the fact that the time needed for the flux change $\phi_{0}$ is the half period $\pi/\omega$. Consequently, as we would expect, the work needed to force the positive charge $e$ across the Josephson junction equals, at least up to a numerical factor, $\hbar \omega$.

Ideally, one would exponentially suppress the energy loss by keeping the tunnel barrier height large by dynamically manipulating the effective critical current $i_{0}$ of the Josephson junction \cite{tunable rfSQUID qubit}.

\subsection{Excitation of quasiparticles}
 Various noise sources contribute to the finite qubit coherence time. For instance, even at sufficiently low temperatures non-equilibrium quasiparticles are present in practice, leading to relaxation of a qubit \cite{quasiparticle relaxation}. Although practical matters are important, here we are content to deal only with fundamental matters. While experimentally observed coherence time for superconducting qubits, on the order of $\mu \mathrm{S}$ \cite{Blais scheme}, may be regarded to be already long enough compared to what we need for the one-shot charged particle detection, we have a new situation here, i.e., the passages of the charged particle.
 
 The excitation of quasiparticles by the passage of a charged particle is something we cannot avoid. This is in contrast with quasiparticle excitations due to infrared radiation from the high temperature parts of the charged particle optics through the entry/exit apertures, because this can in principle be avoided by cooling the entire optics, whether it is practical or not. The problem of excitation by the passing charged particle is especially significant when the charged particle is light, as in the case of the electron, because the light particle tends to move fast. As shown below, a shorter time period leads to more dissipation. For the sake of concreteness, we will talk about electrons in the rest of this subsection.
 
 A fairly rigorous analysis of a flying electron near a superconductor would use the BCS Hamiltonian, with a potential energy term representing an external time-varying electric field, to see whether quasiparticles are excited. Here we perform a much simpler analysis, whose only purpose is to estimate the order of magnitude. Suppose that the rf-SQUID has the dimension of approximately $100\mu \mathrm{m}$ \cite{SQUID qubit experiment}. 
Since electrons with a sufficiently high energy moves approximately at the speed of light, the time scale $\tau$ involved is about 0.3~pS. The energy scale $E/\tau$ turns out to be about 20~meV, which is two orders of magnitude greater than the energy gap $\Delta=180 \, \mu\mathrm{eV}$ of aluminum, the most popular material for the superconducting qubits. Hence we are dealing with an intrinsically high speed phenomena from the perspective of superconductivity. 
It is known that energy dissipation in a superconductor due to alternating current is not much different from that in the normal state, provided that the frequency of the alternating current is above $\Delta/h$. Hence we assume that, in our situation, the rf-SQUID behaves like a normal conductor with a resistance $R$. Roughly, we have the electrical current $e/\tau$, power dissipation $R e^{2}/\tau^{2}$, energy dissipation $\Delta E = Re^{2}/\tau$, and the change of action $\Delta A = R e^{2}=\left( R/R_{Q}\right) h$, where $R_{Q}=h/e^{2} = 25.8 \, \mathrm{k} \Omega$ is the von Klitzing constant. We would expect $\eta  \equiv R/R_{Q}$ to be small compared to 1. Then, first, quantum coherence is essentially protected because $\Delta A \ll h$. Second, the desired condition that energy dissipation being smaller than the energy gap, namely $\Delta E < \Delta$, is expressed as $\eta h/\tau < \Delta$. It might seem paradoxical that the charged particle travels without quasiparticle excitation, i.e., without energy dissipation, when this inequality is satisfied. A plausible answer is that the dissipation occurs {\it on average} as we expect. The probability of quasi-particle excitation is considered small if the inequality is satisfied with much leeway, e.g. $\eta < 10^{-3}$. However, exactly how quasiparticles will be excited and how they affect the qubit operation once these are excited are not clear in the above rough estimate. For example, it could be argued that unless a quasiparticle tunnels through the Josephson junction, essentially nothing significant happens. Furthermore, diffusion of quasiparticles could be blocked by the use of a flux transformer. Hence further investigations are warranted.

\section{Discussion}
\label{sec: discussion}

\subsection{The operation sequence of the rf-SQUID}
\label{operation sequence}
 Among many species of superconducting qubits, the rf-SQUID qubit is far from the easiest to control. In particular, when the two qubit states have the difference in trapped flux close to $\phi_{0}$, as is required in our scheme, the typical barrier height is too high to allow for appreciable tunneling probability between the two state. In fact, the experiment that showed the coherence of rf-SQUID qubit \cite{SQUID qubit experiment} employed excited states in each potential well. An alternative strategy is the use of dynamic modulation of the barrier height \cite{tunable rfSQUID qubit}, but its quantum mechanically coherent use has not been experimentally validated. Whether the use of excited states, or the dynamic modulation of the barrier height is more advantageous remains to be seen at present. For example, the analysis in Sec. \ref{Excitation of the qubit state} needs to be reexamined if we use the excited states.

Another possible avenue for future investigations include the use of multiple superconducting persistent current qubits \cite{persistent current qubit}, despite the fact that each of such qubit has quantum states associated with a much smaller magnetic flux compared to $\phi_{0}$. These devices could be combined by the use of a flux transformer. The use of multiple qubit seems a serious possibility now, given recent demonstrations on controlling multiple superconducting qubit \cite{surface code}.

\subsection{A possible application of the charged particle counter}
 Non-invasive charged particle counting could enable nano-scale assembly of atoms and/or molecules on a substrate, providing these objects can be ionized. The reason is because our scheme is for any single-charged object. Consider an instrument similar to the low energy electron microscope, to which ions are introduced. The rf-SQUID is placed somewhere in the instrument to count the ions going through it. The density of ions are made sufficiently low that only zero or one ions are counted within a suitable time window. The ion is then decelerated at the objective lens of the low energy electron microscope, lands on a substrate, and is electrically neutralized. Repeating this at desired locations on the substrate, one would be able to form an array of atoms/molecules in any desired pattern.

In principle, one could envision to apply this scheme to ionized biological molecules because there are ways to produce large ionized biological molecules \cite{biomolecule ionization}. This might potentially be a useful tool for synthetic biology because one could immobilize the deposited molecule on a cryogenic substrate during the assembly process. For this idea to succeed, however, the landing energy of the biological molecule should be sufficiently low that the molecule would not be damaged, but at the same time minute charging of the substrate should not significantly affect the trajectory of the molecule.

In actual implementations, keeping the rf-SQUID at the dilution-refrigerator temperature while maintaining electron/ion optical access is an issue, since infrared radiation is known to affect superconducting qubit performance \cite{infrared on qubit}. In view of experimental difficulty, however, this scheme does not require coherence of the charged particle waves and hence should be an accessible stepping stone towards realizing entanglement-assisted electron microscopy \cite{eeem, okamoto-nagatani}.

\subsection{Multi-qubit schemes}
 The above single-qubit charged particle detector scheme can be extended to a multiple-qubit area detector version. Since coherence of the charged particle wave is of particular interest here, we will talk about electrons rather than generic charged particles in this subsection in order to be realistic. We will call the electron optical plane of the array of multiple qubits, or any plane conjugate to it, an image plane. The `far field' with respect to the array, or any plane conjugate to it, will be called a diffraction plane. Because of the linearity of quantum mechanics, the extended scheme allows us to transfer the electron quantum state to a quantum memory as shown below. When combined with a quantum information transfer method for the reverse direction, which will also be discussed below, any multi-pixel quantum tasks could be performed. A sensible example of such a task is quantum enhanced multiple phase estimation \cite{Humphreys et al} for low-dose electron microscopy. 

 Consider a 2-dimensional array of $N$ rf-SQUIDs, which we label $0,1,\cdots,N-1$. The initial quantum state of all rf-SQUID qubits are $|s\rangle_{q}$. The state of the rf-SQUID qubit array A, in which only the $k$-th qubit is excited to $|a\rangle_{q}$, is written as $|k\rangle_{A}$. Let the electron state going through the $k$-th rf-SQUID qubit be $|k\rangle$, and the initial (unknown) electron state be $\sum_{k=0}^{N-1}c_{k}|k\rangle$. Because of linearity, after going through the 2D array the state of the system becomes $\sum_{k=0}^{N-1}c_{k}|k\rangle|k\rangle_{A}$. The transmitted electron is then detected in the far field with a conventional single electron area detector. Suppose that the electron is detected in a diffracted state $|D_{s}\rangle=N^{-1/2}\sum_{k=0}^{N-1}e^{i\theta_{k}}|k\rangle$, where $s$ is an integer labeling the detection pixel and phases $\theta_{k}$ are known from the electron optical geometry. Let $|D_{0}\rangle,|D_{1}\rangle,|D_{2}\rangle, \ldots$ be orthonormal basis states. Then, the matrix $A_{n,k}$ that satisfies $|D_{n}\rangle=\sum_{k=0}^{N-1}A_{n,k}|k\rangle$ is unitary. In particular, $A_{s,k}=N^{-1/2}e^{i\theta_{k}}$. Because of the unitarity, we can also write $|k\rangle=\sum_{n=0}^{N-1}A_{n,k}^{*}|D_{n}\rangle$. Since the state before the electron detection can be expanded as
\begin{equation}
\sum_{k=0}^{N-1}c_{k}|k\rangle|k\rangle_{A}=\sum_{k=0}^{N-1}c_{k}A_{s,k}^{*}|D_{s}\rangle|k\rangle_{A}
+\sum_{\substack{n=0 \\ n\neq s}}^{N-1}\sum_{k=0}^{N-1}c_{k}A_{n,k}^{*}|D_{n}\rangle|k\rangle_{A},
\end{equation}
after the electron is detected in the state $|D_{s}\rangle$, the 2D detector is left in the state 
\begin{equation}
\sum_{k=0}^{N-1}c_{k}A_{s,k}^{*}|k\rangle_{A}\propto\sum_{k=0}^{N-1}c_{k}e^{-i\theta_{k}}|k\rangle_{A}.
\end{equation}
 After suitable single-qubit phase manipulations, one will have transferred the electron quantum state to the rf-SQUID qubit array. 

 Transferring quantum information back to an electron is more involved and we assume the availability of a quantum computer. The basic idea
is to use the qudit-version of quantum teleportation \cite{Q teleportation}. The quantum state to be transferred to an electron is prepared in
a register R of the quantum computer as $\sum_{k=0}^{N-1}d_{k}|k\rangle_{R}$.
 First, an electron is generated in a plane wave state, which is $N^{-1/2}\sum_{k=0}^{N-1}|k\rangle$. The electron wave then goes through an rf-SQUID qubit array placed on an image plane, which we assume to be 1-dimensional for now. After interaction, the state of the electron and the rf-SQUID qubit array is $N^{-1/2}\sum_{k=0}^{N-1}|k\rangle|k\rangle_{A}$. The next step is to use the quantum computer to perform a Bell measurement on the combined system of the rf-SQUID qubit array and the resister R. (Meanwhile, the electron may have to go through an electron-optical version of a delay line.)  Specifically, the state is measured with respect to basis states
\begin{equation}
|\psi_{n,m}\rangle_{AR}=\frac{1}{\sqrt{N}}\sum_{k=0}^{N-1}e^{2\pi i\frac{kn}{N}}|k\rangle_{A}|\left(k+m\right)\mathrm{mod}N\rangle_{R},
\end{equation}
where the range of the labels $n$ and $m$ are $0,1,\cdots,N-1$. It is straightforward to check that these states are orthogonal and hence we can write 
\begin{equation}
|\psi_{n,m}\rangle_{AR}=\sum_{k,k'}B_{\left( n,m \right),\left( k,k' \right)}|k\rangle_{A}|k'\rangle_{R},
\end{equation}
where
\begin{equation}
B_{\left( n,m \right),\left( k,k' \right)}=\frac{\delta_{\left(k+m\right)\mathrm{mod}N,k'}e^{2\pi i \frac{kn}{N}}}{\sqrt{N}}
\end{equation}
is unitary. It follows that
\begin{equation}
|k\rangle_{A}|k'\rangle_{R}=\sum_{n,m}B_{\left( n,m \right),\left( k,k' \right)}^{*}|\psi_{n,m}\rangle_{AR}
\end{equation}
and the state of the whole system is
\begin{equation}
\frac{1}{\sqrt{N}}\sum_{k,k'}d_{k'}|k\rangle|k\rangle_{A}|k'\rangle_{R}=\frac{1}{\sqrt{N}}\sum_{k,k',n,m}d_{k'}B_{\left( n,m \right),\left( k,k' \right)}^{*}|k\rangle|\psi_{n,m}\rangle_{AR}
\end{equation}
If the Bell measurement outcome is $\left(n,m\right)$, then the electron state is
\begin{equation}
\sqrt{N}\sum_{k,k'}d_{k'}B_{\left( n,m \right),\left( k,k' \right)}^{*}|k\rangle=\sum_{k'=0}^{N-1}e^{-2\pi i\frac{\left(k'-m\right)n}{N}}d_{k'}|\left(k'-m\right)\mathrm{mod}N\rangle,
\end{equation}
where the overall factor is normalized. Finally, the electron wave is manipulated classically, depending on the outcome $\left(n,m\right)$. The phase factors $e^{-2\pi i\left(k'-m\right)n/N}$ can be compensated for by applying phase shifts pixelwise at the image plane, using e.g. a multi-pixel version of the obstruction-free phase shifter \cite{Rose phase shifter}. We obtain
\begin{equation}
\sum_{k=0}^{N-1}d_{k}|\left(k-m\right)\mathrm{mod}N\rangle.
\end{equation}
Another restoration step $|\left(k-m\right)\mathrm{mod}N\rangle\rightarrow|k\rangle$ can be carried out similarly by another pixelwise phase shifter on a diffraction plane in the electron optical setup. The states $|k\rangle$ can be expressed as
\begin{equation}
|k\rangle=\frac{1}{\sqrt{N}}\sum_{s=0}^{N-1}e^{2\pi i \frac{ks}{N}}|D_{s}\rangle,
\end{equation}
in terms of diffracted states $|D_{s}\rangle$. Hence the electron state is expressed as 
\begin{equation}
\frac{1}{\sqrt{N}}\sum_{k,s}d_{k}e^{2\pi i \frac{\left( k-m \right) s}{N}}|D_{s}\rangle,
\end{equation}
To compensate, we apply a phase shift $2\pi ms/N$ to the electron state $|D_{s}\rangle$. This scheme can be extended to the 2-dimensional case if each pixel is numbered in the raster-scanning manner (See Appendix). In the latter case, each of the two classical phase shifters, located respectively at an image plane and a diffraction plane, consists of two modified obstraction-free
phase shifting devices oriented orthogonal to each other. 

 The above scheme is essentially universal in that anything programmable can be done, including generation of entangled electrons \cite{comment 1}. 

\begin{acknowledgments}
This research was supported in part by the JSPS Kakenhi (grant No. 25390083).
\end{acknowledgments}

\appendix*
\section{The case of a 2-dimensional rf-SQUID qubit array}
 Here we spell out steps for quantum information transfer from the 2-dimensional rf-SQUID qubit array to an electron. 

 We assume that the rf-SQUIDs are on a square lattice on the $xy$ plane, which is perpendicular to the optical $z$-axis. Let the number of rf-SQUIDs along the $x$ and $y$ axes be respectively $N_{x}$ and $N_{y}$, so that the total number of rf-SQUID is $N=N_{x}N_{y}$. Each rf-SQUID has a label $\left(k_{x},k_{y}\right)$ comprising two integers, with the range $0\leq k_{x}<N_{x}$ and $0\leq k_{y}<N_{y}$. These two integers $k_{x},k_{y}$ respectively specifies the position of the rf-SQUID along the $x$ and $y$ axes. Next, we define a single-integer label $k \equiv k_{x}+N_{x}k_{y}$, which range from $0$ to $N-1$. With this label, the argument in the main text goes through without modification also in the present 2-dimensional case.

 However, it may be useful to elaborate on the final classical electron-wave manipulation step. Let us write $k'=\left(k-m\right)\,\mathrm{mod}\, N$.
The electron state before the final step is 
\begin{equation}
\sum_{k=0}^{N-1}e^{-2\pi i \frac{\left(k-m\right)n}{N}}d_{k}|\left(k-m\right)\,\mathrm{mod}\, N\rangle=\sum_{k'=0}^{N-1}e^{-2\pi i \frac{k'n}{N}}d_{k}|k'\rangle
\end{equation}
Note that $d_{k}=d_{\left(k'+m\right)\mathrm{mod}N}$. This can be written as
\begin{equation}
\sum_{k'=0}^{N-1}e^{-2\pi i\frac{n}{N}k_{x}'}e^{-2\pi i\frac{n}{N_{y}}k_{y}'}d_{k}|k'\rangle.
\end{equation}
Hence, the experimenter can first apply a phase shift $2\pi nk_{x}'/N$ to the $k_{x}'$-th row, and then another phase shift $2\pi nk_{y}'/N_{y}$ to the $k_{y}'$-th column to obtain 
\begin{equation}
\sum_{k'=0}^{N-1}d_{k}|k'\rangle.
\end{equation}
 In terms of instrumentation, the first phase shift can be applied with the multi-pixel version of the obstruction-free phase shifter \cite{Rose phase shifter} oriented along the $x$-axis, and the second phase shift can likewise be applied by another one aligned with the $y$-axis.

 To further correct the above state, transformation $|k'\rangle\rightarrow|k\rangle$ is carried out. To do so, we need to go to the Fourier space. This is done naturally in electron optics, as one can use a lens system to obtain the far-field wavefunction. Let us label the pixels in the far field with integers $s=0,1,2,\cdots,N-1$ and write the diffracted electron state going to the $s$-th pixel $|D_{s}\rangle$. Following the numbering method used in the image plane, we write $s=s_{x}+N_{x}s_{y}$, and also $m=m_{x}+N_{x}m_{y}$. 
 Since the transformation between $|k'\rangle$ and $|D_{s}\rangle$ is essentially the 2-dimensional Fourier transform, we write
\begin{equation}
|k'\rangle=\frac{1}{\sqrt{N}}\sum_{s_{x}=0}^{N_{x}-1}\sum_{s_{y}=0}^{N_{y}-1}e^{2\pi i\frac{k_{x}'s_{x}}{N_{x}}}e^{2\pi i\frac{k_{y}'s_{y}}{N_{y}}}|D_{s}\rangle.
\end{equation}
Analogous to the image-plane case, the experimenter can first apply a phase shift $2\pi m_{x}s_{x}/N_{x}$ to the $s_{x}$-th row, and then another phase shift $2\pi m_{y}s_{y}/N_{y}$ to the $s_{y}$-th column in the diffraction plane to obtain $|k\rangle$ back in the image plane. Because of the principle of superposition, this means that the above state $\sum_{k'=0}^{N-1}d_{k}|k'\rangle$ transforms to $\sum_{k=0}^{N-1}d_{k}|k\rangle$, which is what we wanted.


\begin{thebibliography}{20}
\bibitem[25]{comment 1} In a sense, any set of identical fermions are entangled because of antisymmetrization of the wavefunction. Needless to say, this is not what is meant here.
\end{thebibliography}
\end{document}